\documentclass{article}
\usepackage{amssymb}
\usepackage{graphicx}
\usepackage{amsmath}

\input{tcilatex}
\begin{document}

\title{Dark Energy or local acceleration?}
\date{}
\author{Antonio Feoli$^{a}$ and Elmo Benedetto${\,}^{a}$ }
\maketitle

\begin{abstract}
We find that an observer with a suitable acceleration relative to the frame
comoving with the cosmic fluid, in the context of the FRW decelerating
universe, measures the same cosmological redshift as the $\Lambda CDM$
model. The estimated value of this acceleration is $\beta \simeq 1.4\times
10^{-9}m/s^{2}$. The problem of a too high peculiar velocity can be solved
assuming, for the observer, a sort of helical motion.
\end{abstract}

\ {$^{a}$ } Department of Engineering, University of Sannio, Piazza Roma 21,
82100--Benevento, Italy

\section{Introduction}

This work has been motivated by the papers by Tsagas \cite{key-1},\cite%
{key-2},\cite{key-3},\cite{key-4} which discuss the possibility that
peculiar velocities relative to the Hubble flow can simulate a change in the
expansion rate of the universe. For example an observer in Milky Way, that
has a drift flow of $600Km/s$ relative to the comoving cosmological frame,
can measure accelerated expansion within a decelerating universe. In a
recent paper Tsagas \cite{key-5} underlines that the peculiar velocities can
be "well in excess of those anticipated by the current cosmological
paradigm" both due to "dark flows" \cite{key-6},\cite{key-7},\cite{key-8},%
\cite{key-9} and to "bulk flows" \cite{key-10},\cite{key-11},\cite{key-12}
and also that many authors have already considered a possible anisotropy in
the dark energy and the existence of a preferential axis in Cosmic Microwave
Background (CMB). Indeed they assume that the cosmic acceleration is not
uniform in all directions and the universe could expand faster in some
directions than in others \cite{key-13},\cite{key-14},\cite{key-15},\cite%
{key-16},\cite{key-17}. Hence there are a series of discussions and new
proposals about the Dark Energy paradigm. If an alternative scenario is
possible, we want to follow Tsagas' idea but we take a different approach.

The analysis of the motion of a body with a constant proper acceleration in
the x-direction of Minkowski spacetime can be found in many textbooks and it
is called "hyperbolic motion", because its plot in the (x, ct) plane is a
rectangular hyperbola \cite{key-18},\cite{key-19}. This fact was first noted
by Minkowski \cite{key-20} and then by Born \cite{key-21} who also coined
its name, but today the metric of uniformly accelerated observers is known
as Rindler Spacetime. Rindler in fact, in 1960, generalized the concept of
hyperbolic motion to an arbitrary curved spacetime, applied this result to
Schwarzschild and de Sitter spacetimes \cite{key-22} and then studied the
relation with Kruskal Space \cite{key-23}. We want to use his approach to
show how a local acceleration of the observer's reference frame can simulate
some effects of a global acceleration of the expansion of galaxies due to
the action of a dark energy. We underline again that, at this stage, the aim
of the paper is "to simulate some effects of global acceleration" not to
present a complete theory that is alternative to the Dark Energy hypothesis.

\section{Generalized hyperbolic motion}

Summarizing the Rindler approach \cite{key-22}, he considers a particle
having world-line $x^{\mu }(\tau )$ and defines the corresponding velocity $%
U^{\mu }=dx^{\mu }/d\tau $ and acceleration $A^{\mu }$ built through
covariant derivative with respect to the proper time of the accelerated
observer $\tau $ 
\begin{equation}
A^{\mu }=\frac{DU^{\mu }}{d\tau }=\frac{d^{2}x^{\mu }}{d\tau ^{2}}+\Gamma
_{\nu \sigma }^{\mu }\frac{dx^{\nu }}{d\tau }\frac{dx^{\sigma }}{d\tau }
\end{equation}%
Then he studies the dynamics of a uniformly accelerated frame such that 
\begin{equation}
A^{\mu }A_{\mu }=-\alpha ^{2}=constant
\end{equation}%
Rindler shows that the hyperbolic motion in curved spacetime satisfies the
equation 
\begin{equation}
\frac{DA^{\mu }}{d\tau }=\frac{\alpha ^{2}}{c^{2}}U^{\mu }.
\end{equation}%
The above equation has, as a first integral, the solution 
\begin{equation}
U^{\mu }=c\left( \cosh \frac{\alpha \tau }{c}\right) L^{\mu }+c\left( \sinh 
\frac{\alpha \tau }{c}\right) M^{\mu }
\end{equation}%
where the four-vectors $L^{\mu }$ and $M^{\mu }$ are such that $DL^{\mu
}/d\tau =DM^{\mu }/d\tau =0$, $L^{\mu }L_{\mu }=1$, $M^{\mu }M_{\mu }=-1$
and $L^{\mu }M_{\mu }=0$.

In curved spacetime one must find the right expression of these two
four-vectors while in flat spacetime they assume the trivial form $L^{\mu
}=(1,0,0,0)$ and $M^{\mu }=(0,1,0,0)$ and, integrating again, Rindler
obtains 
\begin{equation}
t(\tau )=\frac{c}{\alpha }\sinh \frac{\alpha \tau }{c}
\end{equation}%
and 
\begin{equation}
x(\tau )=\frac{c^{2}}{\alpha }\left( \cosh \frac{\alpha \tau }{c}-1\right)
\end{equation}%
that are the transformations which relate the coordinates (t, x) of the
inertial frame with the proper time $\tau $ of the accelerated observer.

\section{Global or local acceleration?}

The current models of cosmology are based on the following Einstein's
equations 
\begin{equation}
R_{\mu \nu }^{~}-\frac{1}{2}g_{\mu \nu }R=(8\pi G/c^{4})T_{\mu \nu
}^{~}+\Lambda g_{\mu \nu }
\end{equation}%
where $R_{\mu \nu }^{~}$ is the Ricci tensor, $R$ is the Ricci scalar, $%
T_{\mu \nu }^{~}$ is the stress-energy tensor, and $\Lambda $\ is the
cosmological constant. For a standard perfect fluid 
\begin{equation}
T_{\mu \nu }^{~}=\left( p+\epsilon \right) u_{\mu }u_{\nu }-pg_{\mu \nu }
\end{equation}%
where $p$ is the pressure, $\epsilon $ the energy density and $u_{\mu }$ the
velocity of the fluid respectively. Assuming homogeneity and isotropy, the
Friedman-Robertson-Walker (FRW) metric describes the geometry of a spacetime
where the spatial curvature $k=0,\,\pm 1$ is constant 
\begin{equation}
ds^{2}=c^{2}dt^{2}-a^{2}(t)\left[ \frac{dr^{2}}{1-kr^{2}}+r^{2}(d\theta
^{2}+\sin ^{2}\theta d\varphi ^{2})\right]
\end{equation}%
in which $(r,\theta ,\varphi )$ are the comoving coordinates and $a(t)$ is
the scale factor. The corresponding field equations can be written this way 
\begin{equation}
\frac{\overset{..}{a}}{a}=-\frac{4\pi G}{3c^{2}}(\epsilon +3p)+\frac{\Lambda
c^{2}}{3}
\end{equation}

\begin{equation}
\left(\frac{\overset{.}{a}}{a} \right)^{2}+\frac{kc^{2}}{a^{2}}=\frac{8\pi G%
}{3c^{2}}\epsilon +\frac{\Lambda c^{2}}{3}
\end{equation}

\begin{equation}
\overset{.}{\epsilon }+3\left( \frac{\overset{.}{a}}{a}\right) (\epsilon
+p)=0
\end{equation}%
where a dot denotes a derivative with respect to the cosmic time $t$.

It is well known that the simplest FRW model assumes, in the matter
dominated era, zero curvature, zero cosmological constant and zero pressure,
obtaining the following scale factor \cite{key-24},\cite{key-25} 
\begin{equation}
a(t)=a_{0}\left( \frac{t}{t_{0}}\right) ^{2/3}
\end{equation}%
where the subscript $0$ means "at the today time".

At the end of the twentieth century, a series of redshift observations of
distant Type Ia Supernovae \cite{key-26},\cite{key-27} and the study of
Cosmic Microwave Background Radiation, from COBE satellite to the last
results of WMAP and Planck \cite{key-28},\cite{key-29} have shown that the
universe is in an accelerated phase. Therefore it has been necessary to
introduce, in the standard Friedman model, a cosmological constant $\Lambda
\neq 0$. The value of this constant is $\Lambda $ $\simeq 1.1\times
10^{-52}m^{-2}$ as estimated by Planck satellite data of the Cosmic
Background Radiation \cite{key-29}. The resulting $\Lambda CDM$ model is so
used to describe the evolution of the Universe with three fundamental
ingredients: Dark Energy due to the cosmological constant, Cold Dark Matter
and Baryonic Matter. Defining a critical energy density $\epsilon
_{c}=3H_{0}^{2}c^{2}/8\pi G$, the corresponding quantities of Matter and
Dark Energy are divided, following the Planck results, in $\Omega
_{m}=\epsilon _{0}/\epsilon _{c}=0.308$ and $\Omega _{\Lambda }=\Lambda
c^{2}/3H_{0}^{2}=0.692$ while the estimated value of the Hubble constant is $%
H_{0}=67.8Km\cdot s^{-1}\cdot Mpc^{-1}$. In the case $\Lambda \neq 0$, the
solution of equations (10) (11) and (12) for the scale factor is \cite%
{key-30} 
\begin{equation}
a(t)=a_{0}\left( \frac{\Omega _{m}}{\Omega _{\Lambda }}\right) ^{1/3}\sinh
^{2/3}\left( \frac{3}{2}\sqrt{\Omega _{\Lambda }}H_{0}t\right)
\end{equation}%
Note that for $t=t_{0}$ 
\begin{equation}
\left( \frac{\Omega _{m}}{\Omega _{\Lambda }}\right) ^{1/3}\sinh
^{2/3}\left( \frac{3}{2}\sqrt{\Omega _{\Lambda }}H_{0}t_{0}\right) =1
\end{equation}%
from which one can compute the age of the universe 
\begin{equation}
t_{0}=\frac{2}{3H_{0}\sqrt{\Omega _{\Lambda }}}\ln \left( \frac{1+\sqrt{%
\Omega _{\Lambda }}}{\sqrt{\Omega _{m}}}\right)
\end{equation}%
and, using for the constants the estimations of Planck satellite, $%
t_{0}\simeq 13.85$ billions of years.

Now we consider the cosmological redshift 
\begin{equation}
z=\frac{a_{0}}{a}-1
\end{equation}%
From the equation (13), for the FRW model, the redshift is 
\begin{equation}
z_{FRW}=\left( \frac{t_{0}}{t}\right) ^{2/3}-1
\end{equation}%
Using (14) and (15) for $\Lambda CDM$ model the redshift can be expressed
this way 
\begin{equation}
z_{\Lambda CDM}=\left( \frac{\sinh (\frac{3}{2}\sqrt{\Omega _{\Lambda }}%
H_{0}t_{0})}{\sinh (\frac{3}{2}\sqrt{\Omega _{\Lambda }}H_{0}t)}\right)
^{2/3}-1
\end{equation}%
Now we suppose that we are accelerating and that the cosmic time $t$ is
related to the proper time $\tau $ of our accelerated frame by the
transformation (5). If we apply this transformation to the time contained in
the FRW redshift (18) we obtain 
\begin{equation}
z=\left( \frac{\sinh (\frac{\alpha \tau _{0}}{c})}{\sinh (\frac{\alpha \tau 
}{c})}\right) ^{2/3}-1
\end{equation}%
Comparing the equations (19) and (20) we have the same cosmological redshift
as the $\Lambda CDM$ model if 
\begin{equation}
\alpha =\frac{3}{2}\sqrt{\Omega _{\Lambda }}H_{0}c
\end{equation}%
If we were in flat spacetime we would be sure that $\alpha $ represents the
acceleration of our frame with respect to the inertial frame. But now we are
accelerating with respect to the FRW comoving frame and FRW spacetime is
curved, hence the meaning of $\alpha $ is not so trivial. We must explicitly
calculate the covariant acceleration in the FRW background.

\section{Determination of the acceleration}

Our starting point is to preserve the equation (5) and we want to find the
corresponding accelerated motion. From the constraint $U^{\mu }U_{\mu
}=c^{2} $ we obtain 
\begin{equation}
U^{\mu }=c\left[ \cosh \left( \frac{\alpha \tau }{c}\right) ,\pm \frac{\sinh
(\frac{\alpha \tau }{c})}{a(t)},0,0\right]
\end{equation}%
Using the Christoffel symbols of the second kind \cite{key-24} $\Gamma
_{11}^{0}=\frac{a\overset{.}{a}}{c}$and $\Gamma _{01}^{1}=\frac{\overset{.}{a%
}}{ca}$, the components of covariant acceleration are 
\begin{equation}
A^{0}=\frac{DU^{0}}{d\tau }=\alpha \sinh \left( \frac{\alpha \tau }{c}%
\right) +\frac{c\overset{.}{a}}{a}\sinh ^{2}\left( \frac{\alpha \tau }{c}%
\right)
\end{equation}%
\begin{equation}
A^{1}=\frac{DU^{1}}{d\tau }=\pm \left[ \frac{\alpha }{a}\cosh \left( \frac{%
\alpha \tau }{c}\right) +\frac{c\overset{.}{a}}{a^{2}}\sinh \left( \frac{%
\alpha \tau }{c}\right) \cosh \left( \frac{\alpha \tau }{c}\right) \right]
\end{equation}%
from which 
\begin{equation}
A^{\mu }A_{\mu }=-\left[ \alpha +\frac{c\overset{.}{a}}{a}\sinh \left( \frac{%
\alpha \tau }{c}\right) \right] ^{2}=-\beta ^{2}
\end{equation}

From equation (13), using (5), we obtain 
\begin{equation}
\frac{\overset{.}{a}}{a}=\frac{2}{3t}=\frac{2\alpha }{3c\sinh (\frac{\alpha
\tau }{c})}
\end{equation}
So the magnitude of covariant acceleration is 
\begin{equation}
\beta =\alpha +\frac{2}{3}\alpha =\frac{5}{3}\alpha
\end{equation}
hence the motion is uniformly accelerated but the covariant acceleration has
a magnitude $\beta$ different from the corresponding value in the flat case $%
\alpha$. Note also that $A^{\mu }U_{\mu }= 0$ and 
\begin{equation}
\frac{DA^\mu}{d\tau } =\frac{\beta ^{2}}{c^2}U^{\mu }.
\end{equation}%
similar to the equation (3).

Anyway we are able to reproduce the same redshift of $\Lambda CDM$ model
with an accelerating reference frame where 
\begin{equation}
\beta =\frac{5}{2}\sqrt{\Omega _{\Lambda }}H_{0}c
\end{equation}%
Using again the experimental data of Planck collaboration we get $\beta
\simeq 1.37\times 10^{-9}m/s^{2}$. With the slightly different results of
WMAP9 \cite{key-28} $(\Omega _{\Lambda }=0.721,\,\,\,H_{0}=70.0Km\cdot
s^{-1}\cdot Mpc^{-1})$ we have $\beta \simeq 1.45\times 10^{-9}m/s^{2}$.

\section{The problem of peculiar velocity}

In the FRW space-time the physical distance is defined as 
\begin{equation}
D=a(t)r(t)
\end{equation}%
and the corresponding "physical velocity" as 
\begin{equation}
\frac{dD}{dt}=\overset{\cdot }{D}=\overset{\cdot }{a}r+a\overset{\cdot }{r}%
=HD+a\overset{\cdot }{r}
\end{equation}%
This way the velocity with respect to the Hubble flow is 
\begin{equation}
V_{pec}=a(t)\overset{\cdot }{r}(t)
\end{equation}%
and the corresponding peculiar redshift $z_{p}$ is given by the Doppler
formula: 
\begin{equation}
1+z_{p}=\sqrt{\frac{1+(V_{pec}/c)}{1-(V_{pec}/c)}}
\end{equation}%
In our model, from the equation (22), we obtain 
\begin{equation}
V_{pec}=c\tanh \left( \frac{\alpha \tau }{c}\right)
\end{equation}%
Of course, if the acceleration begins at the Big Bang ($t=0$), the
corresponding "peculiar velocity" today ($\tau =t_{0}$) acquires an enormous
value ($V_{pec}\sim 0.8c$), but we do not know when the local acceleration
has started its action (probably during the dust era when structure
formation progresses). So, a possible way to make the value of peculiar
velocity in agreement with experimental data is to take into account a
smaller time window for the action of the acceleration. An alternative
approach is to consider not only a uniformly accelerated observer, but a
more complicated motion, for example a sort of helical trajectory.

We begin writing down the metric (9) for flat spatial curvature ( $k=0$ ) in
the form 
\begin{equation}
ds^{2}=c^{2}dt^{2}-a^{2}(t)\left[ dx^{2}+dy^{2}+dz^{2}\right] .
\end{equation}%
We want to obtain a helical motion, so, if we consider the translational
accelerated motion along the $z$-axis, we can describe the circular
projection on the $xy$ plane by means of the transformations 
\begin{equation}
\left\{ 
\begin{array}{c}
x=R\cos \omega t \\ 
y=R\sin \omega t%
\end{array}%
\right.
\end{equation}%
that, using equation (5), become 
\begin{equation}
\left\{ 
\begin{array}{c}
x=R\cos [\frac{\omega c}{\alpha }\sinh (\frac{\alpha \tau }{c})] \\ 
y=R\sin [\frac{\omega c}{\alpha }\sinh (\frac{\alpha \tau }{c})]%
\end{array}%
\right.
\end{equation}%
Then we obtain the components of the corresponding four-velocity using $%
U^{\alpha }U_{\alpha }=c^{2}$: 
\begin{equation*}
U^{\mu }=\{c\cosh \left( \frac{\alpha \tau }{c}\right) ,-R\omega \sin \left[ 
\frac{\omega c}{\alpha }\sinh \left( \frac{\alpha \tau }{c}\right) \right]
\cosh \left( \frac{\alpha \tau }{c}\right) ,
\end{equation*}%
\begin{equation}
R\omega \cos \left[ \frac{\omega c}{\alpha }\sinh \left( \frac{\alpha \tau }{%
c}\right) \right] \cosh \left( \frac{\alpha \tau }{c}\right) ,\frac{\sqrt{%
c^{2}\sinh ^{2}(\frac{\alpha \tau }{c})-a^{2}R^{2}\omega ^{2}\cosh ^{2}(%
\frac{\alpha \tau }{c})}}{a}\}
\end{equation}%
Starting from this kind of four-velocity, the new components of "physical
velocity" are 
\begin{equation}
\left\{ 
\begin{array}{c}
\overset{\cdot }{D}_{R}=\sqrt{\overset{\cdot }{D}_{x}^{2}+\overset{\cdot }{D}%
_{y}^{2}}=\sqrt{H^{2}+\omega ^{2}}D_{R}=H^{\prime }D_{R} \\ 
where\,\,\,\overset{\cdot }{D}_{y}=\overset{\cdot }{a}y+a\overset{\cdot }{y}%
\,\,\,and\,\,\,D_{R}=aR \\ 
\overset{\cdot }{D}_{z}=\overset{\cdot }{a}z+a\overset{\cdot }{z}=HD_{z}+a%
\overset{\cdot }{z}=H^{\prime }D_{z}+a\overset{\cdot }{z}-(H^{\prime
}-H)D_{z}%
\end{array}%
\right.
\end{equation}%
In this way the "circular motion" can be enclosed in the Hubble flow
considering an effective Hubble constant $H^{\prime }=\sqrt{H^{2}+\omega ^{2}%
}$ while the peculiar velocity with respect to CMB is only due to the
translational motion 
\begin{equation}
V_{pec}=a\frac{d\tau }{dt}\frac{dz}{d\tau }-(H^{\prime }-H)D_{z}=\sqrt{%
c^{2}\tanh ^{2}(\frac{\alpha \tau }{c})-a^{2}R^{2}\omega ^{2}}-(H^{\prime
}-H)az
\end{equation}%
that, choosing suitable constants $R$ and $\omega $, could be made in
agreement with a today value of $V_{pec}\simeq 600Km/s$ \cite{key-5}.
Furthermore, the term under square root of $V_{pec}$ does not become
imaginary in the last ten billions of years.

Our aim to find a local model that is alternative to a global acceleration
has been reached and we do not intend in this paper to study all the details
and consequences of our approach. Of course the covariant acceleration must
be recalculated starting from the new velocity field (38) and the model
could be improved changing assumptions or showing more rigorously the
agreement with the experimental data (fixing, for example, the values of the
arbitrary constants), but we are going to do that in a forthcoming paper.

\section{Conclusions}

We have shown that it is possible to reproduce the behavior of cosmological
redshift in the $\Lambda CDM$ model using the FRW decelerating universe plus
a suitable local acceleration. The task to identify the origin of this
unknown acceleration is beyond the scope of this paper. We observe only that
there are some interesting cases in physics where the acceleration is not
very far from our value such as: the MOND critical acceleration \cite{key-31}%
. $a_{c}\simeq 1.2\times 10^{-10}m/s^{2},$ the Sun's centripetal
acceleration $a_{S}=\left( 220Km/s\right) ^{2}/8Kpc\simeq 1.9\times
10^{-10}m/s^{2}$ and the so called "Pioneer anomaly"\cite{key-32}. that is a
constant acceleration directed towards the Sun of magnitude $a_{p}=\left(
8.74\pm 1.33\right) \times 10^{-10}m/s^{2}$.

As the nature of dark energy is unknown, also the origin of the local
acceleration (if it exists) is still dark.

\section{Acknowledgments}

Thanks to Gaetano Scarpetta for his useful comments. This work was partially
supported by research funds of the University of Sannio.

\end{document}